%% file: main.tex
\documentclass[sigconf,nonacm]{aamas}
\usepackage{balance}
\setcopyright{none}

\usepackage{graphicx}
\usepackage{tabularx}
\usepackage{amsmath,amsthm}
\usepackage{subcaption}
\usepackage[sort&compress,nameinlink,noabbrev,capitalize]{cleveref}
\usepackage{tikz}
\usetikzlibrary{arrows.meta,positioning,calc}
\usepackage{etoolbox}
\usepackage{mathtools}

\usepackage{xcolor}
\usepackage{nicefrac}

\newtheorem{remark}{Remark}
\newtheorem{observation}{Observation}

\newif\ifarxiv
\arxivtrue

\newcommand{\appendixproof}[2]{%
\ifarxiv{}#2\else{}\gappto{\appendixProofText}
{
    \subsection{Proof of \cref{#1}}\label{proof:#1}
    #2
}
\fi{}
}
\newcommand{\appendixProofText}{}

\newcommand{\BibTeX}{\rm B\kern-.05em{\sc i\kern-.025em b}\kern-.08em\TeX}

\title[Grassroots Federation]{Grassroots Federation: Fair Democratic Governance at Scale \\ \smaller{\text{(Full version)}}}

\author{Ehud Shapiro}
\affiliation{%
  \institution{London School of Economics and\\ Weizmann Institute of Science}
  \city{}
   \country{}
}

\author{Nimrod Talmon}
\affiliation{%
  \institution{Ben-Gurion University and \\ Input Output}
  \city{}
  \country{}
}

\begin{abstract}
We propose a framework for the fair democratic governance of federated digital communities that form and evolve dynamically, where small groups self-govern and larger groups are represented by assemblies selected via sortition.  Prior work addressed static fairness conditions; here, we formalize a dynamic setting where federations evolve over time through communities forming, joining, and splitting, in all directions—bottom-up, top-down, and middle-out—and
adapt the fairness guarantees. The main technical challenge is reconciling integral seat allocations with dynamic, overlapping federations, so that child communities always meet their persistent floors while long-run averages converge to proportional fairness. Overcoming these challenges, we introduce a protocol that ensures fair participation and representation both persistently (at all times) and eventually (in the limit after stabilization), extending the static fairness properties to handle structural changes. 

Prior work shows how grassroots federations can be specified via atomic transactions among assembly members,  Constitutional Consensus can realize these transactions and the democratic processes leading to them, and Constitutional Governance in Metric Spaces lets a community govern itself and amend its own constitution. Together, these works form a comprehensive design for an egalitarian, fairly governed, large-scale decentralized sovereign digital community platform.
\end{abstract}

\begin{document}

\maketitle

\input{sections/introduction}
\input{sections/related-work}
\input{sections/model-overview}
\input{sections/static-federations}
\input{sections/grassroots-federations}
\input{sections/fairness-objectives}
\input{sections/greedy-fair-protocol}
\input{sections/analysis}
\input{sections/upward-mobility}
\input{sections/outlook}
\input{sections/acknowledgments}

\bibliographystyle{ACM-Reference-Format}
\bibliography{bib}

\end{document}

%% file: sections/introduction.tex
\section{Introduction}

Social networks have demonstrated that digital communities can scale to billions of members, but under centralized, autocratic control~\cite{zuboff2019age,zuboff2022surveillance}. Decentralized Autonomous Organizations (DAOs)~\cite{ethereum:dao} have shown that digital communities can self-govern, but plutocratically, with governance power proportional to token holdings~\cite{vitalikplutocracy}, while surrendering sovereignty to the operators of the underlying blockchain~\cite{bitcoin,ethereum}. The challenge is the formation and democratic governance of large-scale sovereign digital communities—where members' control does not depend on external authorities/capital.\footnote{A particular acute context is that of blockchain governance; indeed, blockchains face a governance scalability problem: proposals arrive continuously, while human attention is scarce—leading to cognitive load and voter fatigue. Existing approaches mitigate this via delegation and representative layers; our framework is complementary, using federated sortition to keep participation lightweight while retaining legitimacy and diversity. This suggests a concrete design space for on-chain governance: how to sample higher-level assemblies from local ones so that representation remains fair despite hierarchy, overlap, and integer seat constraints.} 

This paper addresses this challenge through a framework for fair democratic governance of communities that form and evolve dynamically.
At the heart of our approach is the concept of \emph{grassroots federation}, which enables democratic governance at scale—from local to global communities the size of today's social networks and potentially all of humanity. The starting point of the grassroots process is the independent formation of small communities—in villages, neighborhoods, professional groups—that govern themselves directly. When communities grow beyond a manageable size for direct governance (say, 100 members), they select a small rotating assembly through sortition to govern on their behalf.
These communities can then federate: A village's dog lovers community joins both the general village federation and the regional dog lovers federation. Villages federate with neighboring villages, professional networks with related guilds. This naturally creates acyclic laminar structures~\cite{halpern2024federated}—from city to state to national assemblies—where communities form based on geography, profession, interests, or identity, with individuals potentially belonging to multiple overlapping communities and communities belonging to more than one parent community. Federations with disjoint populations can coalesce when a person or a community from one joins another, effectively merging the two structures.\footnote{This capability—for federations to coalesce voluntarily while retaining their internal structures—enables bottom-up organization without central coordination. Such federations need not converge to a single apex: a city federation and a professional federation may remain independent, or sovereign communities may coordinate while maintaining separate top-level assemblies. This polycentric architecture—multiple coexisting hierarchies without forced unification—reinforces the grassroots nature of the system, preserving autonomy while enabling coordination.}

Structurally, each community—from a single person to a large collective—is a node in the federation graph, and membership of one community in another is a directed edge in the graph. Each federation is a connected directed acyclic component, and the protocol maintains any number of federations, which may coalesce when a community (including a singleton) from one becomes a child of a community in another.
Communities join parent federations through mutual consent and can leave unilaterally. Each community maintains its own assembly, with parent assemblies drawn from the populations of their children. 
\textbf{The key challenge is to maintain fairness—ensuring each child community has proportional representation in parent assemblies and all individuals equitable chances of selection over time.}\footnote{This follows the long-standing political-theory rationale for sortition and other lotteries in collective decision-making~\cite{stone2011luck}.}

\bigskip
\bigskip

Our work extends Halpern et al.'s foundational work on federated assemblies~\cite{halpern2024federated}, which introduced algorithms for fairly selecting assemblies in a static federation structure. They defined three key fairness properties: individual representation (equal probability of selection for all members), ex ante child representation (expected seats proportional to population), and ex post child representation (actual seats meeting proportional thresholds).\footnote{While Halpern et al.'s model requires parent assembly members to be selected from child assemblies (upward mobility)~\cite{halpern2024federated}, we relax this, allowing parent assemblies to be drawn from the general populations of child communities, enabling fairness guarantees for broader federation structures. In practice, preference for seasoned members can still be incorporated without compromising fairness.}


To extend fairness to dynamic settings, we must reconsider what fairness means when structures continuously evolve. In static settings, fairness is measured at a single point: Does each individual have equal probability of selection? Does each child community receive its proportional share of seats?~\cite{halpern2024federated} In dynamic settings, these become temporal properties. To capture this, we distinguish two kinds of fairness: \textit{persistent fairness}, requiring that certain properties hold at all times—a child community must always have at least its proportional representation (rounded down) in its parent's assembly; and \textit{eventual fairness}, requiring properties to hold in the limit—over time each individual's actual selection frequency should converge to their fair share, and each child community's average representation should converge to its proportional share.\footnote{These eventual fairness properties are defined for the scenario in which the federation structure eventually stabilizes, analogous to the Global Stabilization Time, GST, assumption in distributed systems~\cite{dwork1988consensus}, where protocols are designed to achieve the desired state eventually, once the system stabilizes.
In both cases, this is a theoretical construction, since in practice systems never stabilize forever. Still, it is a well-established practice in theoretical computer science: systems proven to behave well in infinity can then be shown experimentally to exhibit good behaviour in finite horizons in practice.}

The central \textbf{technical challenge} of ensuring both kinds of fairness simultaneously is to guarantee fairness under dynamic, overlapping federations with integral assemblies. Seat entitlements are fractional and change as communities join, leave, or overlap, yet assemblies must always satisfy integer constraints. Naïve strategies---such as simple rounding or uniform rotation---fail: they can violate persistent lower bounds for children or prevent long-run averages from converging to proportional shares. The difficulty is to reconcile these conflicting requirements---persistent floors at every instant and eventual proportionality for both communities and individuals---within one consistent protocol.

\begin{figure*}
\begin{center}
   \includegraphics[width=12cm]{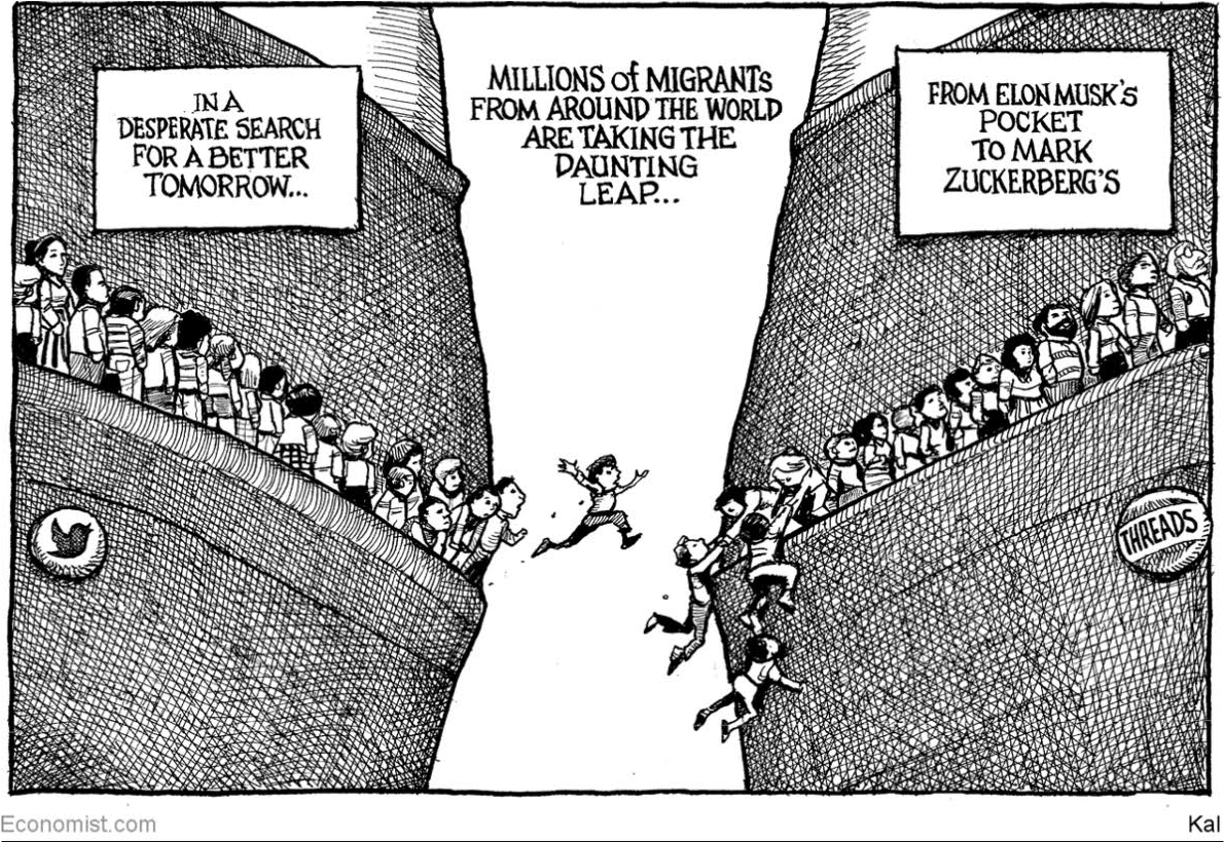}
  \caption{Examples of non-democratic, large-scale, centralized, non-sovereign digital communities. (The Economist, July 13, 2023)  
  }
\label{figure:economist}
\end{center}
\end{figure*}

Grassroots Federation aims to address \textbf{the egalitarian formation and the fair democratic governance of large-scale, decentralized, sovereign digital communities}.  As such, it may offer the first concrete proposal to realize \emph{global democracy} (aka \emph{cosmopolitan democracy})~\cite{archibugi2004cosmopolitan} since envisioned more than two centuries ago by Kant~\cite{kant1795perpetual}.  After all, social networks have shown that a single digital platform can bring billions of people together; the remaining challenge is to build one that does so  democratically.  Developing the conceptual and mathematical foundations for doing that is the subject of this paper

Sortition can ensure that an assembly represents a given strata (e.g.\ female/male, rural/urban, young/old) perfectly in expectation, given a population~\cite{FGGP20,EKMP+22,Bour13,FKP21,flanigan2021fair}.  But the two ``given'' presume a \emph{giver}---an authority that determines both the population and the strata to be represented. How is this authority itself elected democratically? Grassroots Federation resolves this circularity by letting the people themselves determine both: by freely assembling and federating, they establish the populations from which assemblies are selected and the strata to be represented, independently of any external authority.

A simpler `direct-democracy' instance of Grassroots Federation, which captures the grassroots formation of communities and their federation but abstracts away the selection of assemblies, was formally defined and proven grassroots in previous work~\cite{shapiro2025atomic,lewis2026volitional}.  There, joining or leaving a federation is realized by an atomic transaction in which all members of both communities participate, serving as an abstraction of direct digital democracy but requiring untenable large-scale consensus and voting.  The present work replaces direct participation with sortition-based assemblies, and focuses on the fairness of representation and equity of participation that assembly selection, rotation, and maintenance must satisfy in a dynamic setting. It is a follow-on on previous work~\cite{halpern2024federated} that examined fairness in the static setting of a given federation structure. Grassroots Federation specifies how assemblies are selected; Constitutional Governance in Metric Spaces~\cite{shapiro2026cgms} provides a decision rule by which a selected assembly---or, in a small community, the members directly---can govern itself.  

\paragraph{Our contribution.}
We make the following contributions:
(1) We \textbf{formalize} a dynamic model of federation formation, where agents can form overlapping communities that grow and federate over time, governed by sortition-based assemblies; 
(2) We \textbf{define} fairness and representativeness conditions adapted to the dynamic setting—persistent properties, to hold at all times, and eventual properties, to hold in the limit;
(3) We \textbf{design} an efficient protocol for maintaining assemblies across dynamically-evolving communities, proving that it satisfies all fairness guarantees.

\paragraph{Architectural context}
This work provides a building block for realizing the grassroots vision of replacing autocratic and plutocratic global platforms with egalitarian and democratic grassroots platforms~\cite{shapiro2023grassrootsBA,shapiro2025atomic,lewis2026volitional}. To serve as viable alternatives used by billions, grassroots platforms must support democratic governance at comparable scales. Since they consist of multiple, independent instances rather than a single global one, federation is essential for large-scale governance~\cite{shapiro2023grassrootsBA}. Alongside prior work on static fairness~\cite{halpern2024federated}, volitional multiagent atomic transactions~\cite{shapiro2025atomic,lewis2026volitional}, and constitutional consensus~\cite{keidar2025constitutional}, this work completes the architecture needed for democratic governance of grassroots sovereign digital communities—at scale. It ensures fairness properties persist and converge as federations grow.

Technically, this work forms the top layer of a three-layer architecture for democratic governance of grassroots digital communities. 
At the lowest, \textbf{consensus layer}, the \textit{Constitutional Consensus} protocol~\cite{keidar2025constitutional} ensures that community members' personal devices (e.g., smartphones) reach agreement on shared state updates, even as membership changes over time. 
Above it, the \textbf{volitional multiagent atomic transactions layer}~\cite{shapiro2025atomic,lewis2026volitional} specifies how such state changes are structured and executed: each community’s rules—its ``digital social contract''—are represented as volitional atomic transactions among its members, ensuring every valid community update is agreed upon and applied consistently by all participants. 
At the top, the \textbf{federation layer}—the focus of the present paper—governs how communities relate to one another: how assemblies are formed, how representation is maintained, and how fairness persists as communities evolve.  
Implemented together, these layers yield a \emph{grassroots} system~\cite{shapiro2023grassrootsBA,shapiro2025atomic,lewis2026volitional}: one where communities operate and interconnect directly on their members’ devices, without dependence on external authorities or centralized infrastructure.  
This work completes the architecture by providing the fairness guarantees required for democratic governance at scale.\footnote{Note that the intended scope is not merely abstract: it covers (i) human collectives, (ii) fully autonomous multi-agent systems, and (iii) hybrid environments in which AI agents act as constrained proxies for individuals. In all three, the mechanism can be read as a decentralized coordination layer that composes many local decisions into a coherent outcome while keeping participation and communication demands bounded; in mixed settings, this interpretation naturally foregrounds issues of mandate specification and auditability of proxy behavior.}

\newcommand{\People}{P}                       
\newcommand{\Assembly}[1]{A_{#1}}            
\newcommand{\Members}[1]{M_{#1}}             
\newcommand{\Pop}[1]{P_{#1}}                 
\newcommand{\Children}[2][]{\mathrm{children}_{#1}(#2)} 

\newcommand{\wrelperson}[3]{w_{\mathrm{rel}}(#1,#2,#3)}    
\newcommand{\wrelchild}[2]{w_{\mathrm{rel}}(#1,#2)}        
\newcommand{\wpropperson}[3]{w(#1,#2,#3)}                  
\newcommand{\wpropchild}[2]{w(#1,#2)}                      
\newcommand{\repperson}[2]{r(#1,#2)}                       
\newcommand{\repchild}[2]{r(#1,#2)}                        
\newcommand{\mean}[1]{\overline{#1}}                       

\newcommand{\PFR}{Persistently-Fair Representation}   
\newcommand{\EFR}{Eventually-Fair Representation}   
\newcommand{\EEP}{Eventually-Equitable Participation}  

\paragraph{Paper structure.}
After discussing related work (Section~\ref{section:related work}), Section~\ref{section:model} provides a model overview with intuitive examples. Section~\ref{section:static} presents static federations, defining validity conditions and assemblies. Section~\ref{section:dynamic} introduces grassroots federations as a timed transition system, specifying how federations evolve through communities and individuals joining, leaving, and reorganizing. Section~\ref{section:fairness} adapts fairness requirements from the static to dynamic setting, defining persistent and eventual fairness properties. Section~\ref{section:protocol} presents the Greedy Fair Protocol (GFP). Section~\ref{section:analysis} proves that GFP satisfies all fairness objectives. Section~\ref{section:pragmatics} considers extending GFP with upward mobility. Section~\ref{section:outlook} discusses future directions.

An abridged version of this paper appeared in Proc. AAMAS 2026.

%% file: sections/related-work.tex
\section{Related Work}\label{section:related work}

Our work lies at the intersection of large-scale democratic design, apportionment, sortition, and temporal fairness. The main novelty is in supporting \emph{dynamic, overlapping federated assemblies} while maintaining both persistent guarantees and eventual proportional fairness. Prior work typically addresses one of these aspects in isolation: apportionment in static settings, sortition for single draws, or temporal fairness without indivisible seats. We bring these threads together to scale democracy in grassroots federations.

\emph{Apportionment and Federated Assemblies.}  
The classical apportionment literature studies the allocation of indivisible seats proportionally in static settings \cite{balinski2010fair}. Halpern et al.\ recently introduced a model of static federated assemblies, selecting parent representatives from child populations while satisfying fairness constraints \cite{halpern2024federated}. We generalize this setting to dynamic processes with turnover and overlapping memberships.

\emph{Sortition and Randomized Committee Selection.}  
Sortition has been revisited in social choice. Flanigan et al.\ propose a transparent sortition method that achieves fairness under quotas \cite{FKP21}. Baharav and Flanigan extend this with manipulation-robust transparent sortition \cite{baharav2024fair}. We adopt sortition as our base mechanism but embed it into long-lived rotating assemblies under evolving membership.

\emph{Temporal and Dynamic Social Choice.}  
Work on temporal social choice studies how decisions evolve over time \cite{parkes2013dynamic, lackner2020perpetual}. These models examine changing preferences and outcomes, but not proportional representation across assemblies. In contrast, we focus on the dynamics of seat allocation and fairness guarantees in federations.

\emph{Overlapping Communities and Multiwinner Representation.}  
Research on multiwinner voting and proportional representation (e.g., justified representation, proportional justified representation \cite{aziz2017justified}) assumes disjoint electorates. In our setting, populations may overlap across communities, requiring new mechanisms (colored seats) to handle weight-splitting and avoid double-counting.

\emph{Stabilization and Convergence.}  
Our notion of eventual fairness after stabilization is inspired by concepts from distributed systems, particularly Global Stabilization Time (GST) \cite{dwork1988consensus}. As in eventual synchrony, our guarantees hold once the system stops changing, though the application here is democratic seat allocation.

\emph{Democracy at Scale.}  
Efforts to scale democracy include liquid democracy \cite{blum2016liquid}, DAOs, and digital platforms, typically based on delegation or token voting. Grassroots federation is a new paradigm: egalitarian, sortition-based, with provable fairness guarantees.

%% file: sections/model-overview.tex
\section{Model Overview}\label{section:model}

We first sketch the setting informally, to make the core objects and terminology concrete; the following sections (Sections~\ref{section:static} and~\ref{section:dynamic}) will then make
them fully precise.
The central model entity is a \emph{federation}: a network of \emph{communities} that is typically a directed acyclic graph (DAG). Each community governs itself through an \emph{assembly}, a small group of \emph{members} chosen and rotated by \emph{sortition}, meaning members are randomly drawn from the community, where a person can be in several communities and in several assemblies.

Crucially, federations are dynamic: communities can join under a new parent community, which has its own assembly,  drawn from the populations of its children. Communities can also leave and individual people can join or depart communities.
Moreover, members do not serve indefinitely in the assemblies: each member of an assembly is selected for a fixed \emph{term} $\tau$ (say, one year), after which their seat is re-assigned. Term limits ensure regular turnover, giving more members of the community an opportunity to participate, and preventing
long-term concentration of influence in the hands of a few.
This also relates to the following three intuitive fairness goals we aim to capture (note that the first two goals consider child communities, while the last goal considers individual persons):
\begin{itemize}
\item \textbf{Persistent fairness:}
At every moment, each child community should hold at least its guaranteed minimum number of seats in the parent assembly, proportional to its population share.
E.g., a child holding two-thirds of a parent's population of \(100\) is entitled to \(\nicefrac{2}{3} \cdot 100 = 66.66\ldots\) seats, guaranteeing at least \(66\).
\item \textbf{Eventual fairness:}
Over the long run, as seats rotate due to term limits, each child community's time-averaged share of seats should be at least its proportional population share.
In the same example, the community should hold, on average, at least \(66.666\ldots\) seats (the exact real number, not its floor).
\item \textbf{Equitable participation:}
Over the long run, each \emph{person} in a community should serve in its assembly for a similar fraction of time. This ensures no individual is persistently overlooked while others over-participate.
\end{itemize}

These fairness properties interact: persistent fairness prevents sharp drops in representation, while eventual fairness and equitable participation ensure that the system converges to proportional representation for communities and individuals in the long term. Consider Figure~\ref{figure:intuitive example}.
The rest of this section builds the formal model needed to capture such structures, their evolution over time, and the fairness properties we want to enforce.

\begin{figure*}[t]
\centering

\tikzset{
  >=Latex,
  community/.style={
    draw,
    rounded corners=5pt,
    fill=black!2,
    text width=3.05cm,
    minimum height=0.95cm,
    align=center,
    inner sep=2.5pt
  },
  edge/.style={-Latex, thick, draw=black!70, shorten <=2pt, shorten >=2pt},
}

\newcommand{\comm}[2]{%
  \begin{tabular}{@{}c@{}}
    \textit{Community #1}\\[-1pt]
    \scriptsize(#2)
  \end{tabular}%
}
\newcommand{\commAsm}[2]{%
  \begin{tabular}{@{}c@{}}
    \textit{Community #1}\\[-1pt]
    \scriptsize(#2)\\[-2pt]
    \rule{1.65cm}{0.3pt}\\[-3pt]
    \tiny\bfseries Assembly of #1
  \end{tabular}%
}

\begin{subfigure}{0.48\linewidth}
\centering
\begin{tikzpicture}[font=\footnotesize, scale=0.92, every node/.style={transform shape}]
  \node[community, text width=1.4cm] (A) at (0.85,0) {\comm{A}{Alice}};
  \node[community, text width=1.4cm] (B) at (2.55,0) {\comm{B}{Bob}};
  \node[community] (X) at (1.675,1.25) {\commAsm{X}{Alice and Bob}};

  \draw[edge] (X.south) -- (A.north);
  \draw[edge] (X.south) -- (B.north);
\end{tikzpicture}
\caption{Alice \& Bob federate into X.}
\end{subfigure}\hfill
\begin{subfigure}{0.48\linewidth}
\centering
\begin{tikzpicture}[font=\footnotesize, scale=0.92, every node/.style={transform shape}]
  \node[community] (Xc) at (0,-0.2) {\commAsm{X}{Alice and Bob}};
  \node[community] (Cc) at (3.35,-0.2) {\comm{C}{Carol}};

  \node[community] (Y)  at (1.675,1.10) {\commAsm{Y}{Alice, Bob, Carol}};

  \node[community, text width=1.4cm] (A2) at (-0.85,-1.45) {\comm{A}{Alice}};
  \node[community, text width=1.4cm] (B2) at (0.85,-1.45) {\comm{B}{Bob}};

  \draw[edge] (Y.south) -- (Xc.north);
  \draw[edge] (Y.south) -- (Cc.north);

  \draw[edge] (Xc.south) -- (A2.north);
  \draw[edge] (Xc.south) -- (B2.north);
\end{tikzpicture}
\caption{X (built from A and B) federates with Carol into Y.}
\end{subfigure}

\caption{Running example of federating communities with a fixed assembly size $n=2$.
(a) Communities $A$ (containing Alice) and $B$ (containing Bob) federate into a parent community $X$. Each child community has the same size, so $X$’s two assembly seats are split evenly.
(b) The previously formed community $X$ (population~2) federates with $C$ (containing Carol, population~1) into parent community $Y$. Fairness considerations imply that: (1) persistently, each receives at least one seat; and (2) over time, $X$ gets $\approx 1.33$ seats and $C$ $\approx 0.66$ seats.}

\Description{Two-panel diagram showing a small federation. In panel (a), communities A and B are children
of parent X; “Assembly of X” appears inside X’s node. In panel (b), Y has children X and C, and X’s
internal children A and B are shown beneath it; “Assembly of X” and “Assembly of Y” appear inside
their respective nodes.}

\label{figure:intuitive example}
\end{figure*}

%% file: sections/static-federations.tex
\section{Static Federations}\label{section:static}

First, we provide some definitions for the static case.
Let $\Pi$ denote a (potentially infinite) universe of people.
%
Fix an integer $n \ge 1$ as the target assembly size
for every community.

\begin{definition}[Communities and federation graph]
A \emph{federation graph} is a directed acyclic graph $G=(V,E)$ whose nodes $V$ are \emph{communities}. The graph may consist of multiple connected components, each being a \emph{federation}.
For $v \in V$, let
\(
\Children{v} \;:=\; \{\,u \in V : (v \to u) \in E\,\}
\)
denote the set of (immediate) children of $v$. A node is a \emph{leaf} if $\Children{v}=\emptyset$.
\end{definition}

\begin{definition}[Population]
The \emph{population} of a community $v$, denoted $\Pop{v}$, is the set of all people
who belong to leaves in the subtree rooted at $v$ (which is $v$ itself if it is a leaf).
\end{definition}

Each community $v \in V$ is governed through an \emph{assembly} 
denoted $\Assembly{v}$.  
The assembly consists of $n$ members of the population of $v$, or all of them if it's size is $\le n$.  

\begin{definition}[Valid federation]\label{definition:valid federation}
A federation with communities~$V$ and assemblies $\{\Assembly{v} : v \in V\}$ is \emph{valid} if:
\begin{enumerate}
    \item \textbf{Leaf disjointness:}  
    For distinct leaves $u$ and $v$, $\Pop{u} \cap \Pop{v} = \emptyset$.
    \item \textbf{Assembly membership and size:}  
    For every $v \in V$, $\Assembly{v} \subseteq \Pop{v}$ and $|\Assembly{v}| = \min(|\Pop{v}|, n)$.
\end{enumerate}

A federation graph is \emph{valid} if each of its federations is valid.
\end{definition}

We note that in a grassroots setting, the due diligence communities perform on each other before deciding to form an edge among them may include verifying that each other's federation is valid, in case they belong to different federations.

\begin{example}[Valid federation from Figure~\ref{figure:intuitive example}]
Consider the federation in panel~(b) of Figure~\ref{figure:intuitive example}, with target assembly size $n=2$.

\begin{itemize}
    \item The leaves are communities $A$, $B$, and $C$, with populations
    $\Pop{A} = \{\text{Alice}\}$, $\Pop{B} = \{\text{Bob}\}$, and
    $\Pop{C} = \{\text{Carol}\}$, which are pairwise disjoint.
    
    \item Internal community $X$ has population
    $\Pop{X} = \{\text{Alice}, \text{Bob}\}$,
    and its assembly $\Assembly{X}$ contains $n=2$ distinct members from~$\Pop{X}$.

    \item Parent community $Y$ has population
    $\Pop{Y} = \{\text{Alice}, \text{Bob}, \text{Carol}\}$,
    with assembly $\Assembly{Y}$ containing $n=2$ distinct members from~$\Pop{Y}$.
\end{itemize}

Validity holds: (i) leaf populations are disjoint, and (ii) each assembly contains $\min(|\Pop{v}|,n)$ distinct members from its population.
\end{example}

%% file: sections/grassroots-federations.tex
\section{Grassroots Federations}\label{section:dynamic}

The static model above captures a single snapshot of a federation graph: its communities, their populations, and the current assemblies. As our focus is on the grassroots formation of federations, we consider how federations evolve:
\begin{enumerate}
    \item a person may participate in a federation by first forming a singleton community, and join a parent community as a child;
    \item communities can federate by creating parent communities, join existing parents through mutual consent, or leave;
    \item federation graphs evolve as these operations compound over time, potentially resulting in complex DAG structures; and
    \item assemblies continuously evolve to fairly reflect communities joining and leaving, with members rotating via term limits to maintain fair representation over time.
\end{enumerate}

We provide a formal model of such dynamics, by formulating a \emph{timed transition system} whose states are valid federations and whose transitions correspond to the federation events: participation, federation, joining, leaving, and assembly maintenance.

\subsubsection{Timed Transition Systems}

Concretely, to model changes over time, we build on a standard abstraction from concurrency theory.

\begin{definition}[Timed Transition System]
A \emph{timed transition system} is a tuple $\mathcal{T} = (S, s_0, T)$ where:
(1) $S$ is the set of possible states;
(2) $s_0 \in S$ is the initial state;
(3) $T \subseteq S \times S$ is the set of \emph{valid transitions}.

A \emph{timed run} of $\mathcal{T}$ is a sequence
\(
(s_0, t_0) \rightarrow (s_1, t_1) \rightarrow (s_2, t_2) \rightarrow \cdots
\)
such that:
\begin{enumerate}
    \item $(s_i, s_{i+1}) \in T$ for all $i$ (valid transition).
    \item $t_0 = 0$ and $t_{i+1} \ge t_i$ for all $i$ (non-decreasing global clock).
\end{enumerate}
Here $t_i$ is the \emph{absolute time} at which state $s_i$ is current.
\end{definition}


\subsection{Federation Transition System}\label{section:federation transition system}

We now tailor the abstract timed transition system from above to the federation setting. In what follows, we specify the states and event types for this specialized system, which together form our dynamic model of federations.

\begin{definition}[Federation Transition System]\label{definition:fts}
A \emph{federation transition system} is a timed transition system $\mathcal{F} = (\mathcal{G}, G_0, T)$ where:
\begin{itemize}
    \item $\mathcal{G}$ is the set of all \emph{valid federation graphs} (as per Definition~\ref{definition:valid federation}).
    \item $G_0 = (\emptyset,\emptyset)$ is the empty federation (no people, no nodes).
    \item To specify $T \subseteq \mathcal{G} \times \mathcal{G}$, it is useful to first classify the possible transitions/events into the following types:
    \begin{enumerate}
    
        \item \textbf{Participate} $(p)$: Add a new leaf $v$ with $\Pop{v}=\{p\}$ and $\Assembly{v}=\{p\}$, provided $p$ is not in any existing leaf population.
    
        \item \textbf{Federate} $(v)$: Add new node $f$, set $\Assembly{f} := \Assembly{v}$, and add edge $f \to v$. (This event creates a federation ``placeholder'' as a parent of $v$, having $v$ as its sole child, ready to be joined with other children.)
        
        \item \textbf{Join} $(f \to v)$: Add edge $f \to v$ if $f,v \in V$ and the result is acyclic. (I.e., this adds a new child, $v$, to $f$.)
                
        \item \textbf{Leave} $(f \to v)$:
        Remove edge $f \to v$. 
        Remove from $\Assembly{f}$ only those members $p$ who no longer belong to any remaining child of $f$, 
        i.e., $p \notin \bigcup_{u \in \Children[t]{f}} P_u$. 
        (With the colored-seat formalism introduced later, this corresponds to removing the seats colored by $v$.)
        
        \item \textbf{Remove Member} $(p,f)$: Remove $p$ from $\Assembly{f}$. (This corresponds to the end of term for member $p$ in $\Assembly{f}$.)
        
        \item \textbf{Add Member} $(p,f)$: If $p \in \Pop{f} \setminus \Assembly{f}$ and $|\Assembly{f}| < \min(|\Pop{f}|,n)$, add $p$ to $\Assembly{f}$.
        
        \item \textbf{Garbage Collect} $(v)$: If $v$ is a leaf with $\Assembly{v}=\emptyset$, remove $v$ and all incident edges; update parent assemblies accordingly.
    
    \end{enumerate}
    
    \textbf{Given these event types}, $(G, G') \in T$ iff either:
    $G'$ is created from $G$ following an event of type $6$ (\emph{Add Member}) or $7$ (\emph{Garbage Collect}), \textbf{or} no event of type $6$ or $7$ is applicable to $G$, and $G'$ is obtained from $G$ by applying a transition of some other type.
    
\end{itemize}
\end{definition}

In other words, events of type $6$ and $7$ have \emph{priority}: if at least one such event is applicable in $G$, then the only outgoing transitions from $G$ in $T$ are those events.
Furthermore, we require that, in a valid run of our transition system, any event of type (6) or (7) occurs without advancing time; that is, if $(G_i, t_i) \rightarrow (G_{i+1}, t_{i+1})$ and this corresponds to transition of type (6) or (7), then $t_{i+1} = t_i$. Such maintenance events must be applied exhaustively—repeatedly, at the same timestamp—until no further type (6) or (7) event is applicable before any other event type may occur.

\subsection{Validity of the Federation Transition System}

Our first goal (a sanity check) is to show that the federation transition system produces only \emph{valid} federations. Yet not all states in a run persist: some exist only instantaneously as part of system maintenance.
Thus, we distinguish two kinds of states:
\begin{itemize}
    \item \textbf{Non-fleeting states}: states that persist for a positive duration, i.e., $t_i > t_{i-1}$ or the initial state $i=0$.
    \item \textbf{Fleeting states}: arise only for zero time, i.e., $t_i = t_{i-1}$.
\end{itemize}

Only the non-fleeting states correspond to federations that persist ``in the real world''.
The following proposition states that all such non-fleeting states are indeed valid.

\begin{proposition}[Validity of non-fleeting states]\label{prop:valid-observable}
In any admissible timed run $(G_0,t_0)\rightarrow(G_1,t_1)\rightarrow\cdots$ of the federation transition system, every non-fleeting state $G_i$ is a valid federation graph: $G_i \in \mathcal{G}$.
\end{proposition}

\appendixproof{prop:valid-observable}{
\begin{proof}
Consider a run and fix the strictly increasing sequence of non-fleeting indices $0=i_0<i_1<i_2<\cdots$ (if finite, take the maximal one). We prove by induction on $k$ that $G_{i_k}\in\mathcal{G}$. For the \emph{base}, note that $G_{i_0} = G_0 = (\emptyset, \emptyset) \in \mathcal{G}$ by definition.

For the inductive \emph{step}, assume $G_{i_k}\in\mathcal{G}$. By admissibility, $G_{i_k}$ has no applicable maintenance event (otherwise time would not have advanced to $t_{i_k}$), so the next event that
advances time is a single non-maintenance event $e\in\{1,\dots,5\}$ at time $\tau:=t_{i_{k+1}}$.
Let $H$ be the immediate post-event state (before any maintenance events is applied on it). We analyze $e$:
\begin{itemize}
\item \textbf{(1) Participate $(p)$:} Adds a fresh leaf $v$ with $\Pop{v}=\Assembly{v}=\{p\}$.
The precondition “$p$ not in any existing leaf” preserves leaf disjointness; other leaves and
assemblies are unchanged. Hence $H$ is already valid.

\item \textbf{(2) Federate $(v)$:} Creates a new parent $f$ with edge $f\to v$ and
$\Assembly{f}:=\Assembly{v}$. Since $\Pop{f}=\Pop{v}$, we have
$\Assembly{f}\subseteq \Pop{f}$ and $|\Assembly{f}|=\min(|\Pop{f}|,n)$. Leaves unchanged.
Thus $H$ is valid.

\item \textbf{(3) Join $(f\to v)$:} Adds edge $f\to v$ (acyclic by precondition), so
$\Pop{f}$ increases to $\Pop{f}\cup \Pop{v}$ while $\Assembly{f}$ is unchanged.
Membership $\Assembly{f}\subseteq \Pop{f}$ is preserved (population expands).
Leaf disjointness unchanged. The only potential violation is that
$|\Assembly{f}|$ may be \emph{smaller} than $\min(|\Pop{f}|,n)$; zero-time maintenance (6)
fills up to the bound.

\item \textbf{(4) Leave $(f\to v)$:} Removes edge $f\to v$; $\Pop{f}$ shrinks. By the event’s
postcondition we drop from $\Assembly{f}$ any $p\notin \Pop{f}$ (so membership holds).
Leaf sets are unchanged. Now $|\Assembly{f}|\le \min(|\Pop{f}|,n)$ automatically
($|\Assembly{f}|\le |\Pop{f}|$ and it never exceeds $n$), but it may be \emph{smaller}
than the bound after dropping members; maintenance (6) re-fills if needed.

\item \textbf{(5) Remove Member $(p,f)$:} Decreases $|\Assembly{f}|$ by $1$.
Leaf disjointness and membership are unaffected. If $f$ is a leaf and becomes empty,
maintenance (7) removes the leaf; otherwise maintenance (6) may re-fill.
\end{itemize}

Thus, after $e$ at time $\tau$, the only possible invalidities are exactly those repaired by the
zero-time maintenance burst (6)–(7), which is applied exhaustively by admissibility. The resulting
observable state $G_{i_{k+1}}$ (the first state after time $\tau$) satisfies both validity clauses,
so $G_{i_{k+1}}\in\mathcal{G}$.
\end{proof}
}

\subsection{Time-Indexed Notation}

Following the definitions above, we also introduce time-indexed notation, to be able to speak about the state of the federation at some time $t$.
Formally, fix an admissible timed run
\(
(G_0,t_0)\rightarrow(G_1,t_1)\rightarrow(G_2,t_2)\rightarrow\cdots
\)
as in Section~\ref{section:federation transition system}, with $t_0=0$ and $t_{i+1}\ge t_i$.
For absolute time $t\ge 0$, let $i(t) := \max\{\, i \mid t_i \le t \,\}$ and $G_t := G_{\,i(t)}$ be the state that is \emph{current at time $t$} (right-continuous selection; also, recall that maintenance events (types~6 and~7) do not advance time, so, following Proposition~\ref{prop:valid-observable}, every $G_t$ is a valid federation).

For any community $f$ present in $G_t$, we write:
\begin{itemize}
    \item $P_{t,f}$ for the population of $f$ in $G_t$ (at time $t$),
    \item $M_{t,f}$ for the members of the assembly of $f$ in $G_t$ (at time $t$),
    \item $\Children[t]{f}$ for the set of children of $f$ in $G_t$ (at time $t$).
\end{itemize}

\subsection{The Running Example (Dynamic)}

Consider the following example, illustrating how the valid federation of Figure~\ref{figure:intuitive example} could have emerged dynamically.

\begin{example}[Building the running example via events]
Fix $n=2$. We start from $G_0=(\emptyset,\emptyset)$ and perform the following admissible run.
We write $(\mathrm{M})$ for zero-time maintenance (events~6/7) that may occur immediately after an event.

\begin{enumerate}
\item[$t=1$] \textsc{Participate}(\text{Alice}).  
Create leaf $A$ with $\Pop{A}=\{ \text{Alice} \}$ and $\Assembly{A}=\{ \text{Alice} \}$.  
(No maintenance needed.)

\item[$t=2$] \textsc{Participate}(\text{Bob}).  
Create leaf $B$ with $\Pop{B}=\{ \text{Bob} \}$ and $\Assembly{B}=\{ \text{Bob} \}$.  
(No maintenance.)

\item[$t=3$] \textsc{Federate}($A$).  
Create parent $X$ with edge $X \to A$ and set $\Assembly{X} := \Assembly{A} = \{ \text{Alice} \}$.  
Here $\Pop{X}=\Pop{A}=\{\text{Alice}\}$, so $|\Assembly{X}|=\min(|\Pop{X}|,2)=1$. (No maintenance.)

\item[$t=4$] \textsc{Join}($X \to B$).  
Add edge $X \to B$, so $\Pop{X}$ becomes $\{\text{Alice},\text{Bob}\}$ while $\Assembly{X}$ is still $\{\text{Alice}\}$.  
\emph{(M)} Event~(6) \textsc{AddMember} fills $\Assembly{X}$ up to $\min(|\Pop{X}|,2)=2$, adding Bob at the \emph{same} timestamp:  
$\Assembly{X}=\{\text{Alice},\text{Bob}\}$.

\item[$t=5$] \textsc{Participate}(\text{Carol}).  
Create leaf $C$ with $\Pop{C}=\{\text{Carol}\}$ and $\Assembly{C}=\{\text{Carol}\}$.  
(No maintenance.)

\item[$t=6$] \textsc{Federate}($X$).  
Create parent $Y$ with edge $Y \to X$ and set $\Assembly{Y}:=\Assembly{X}=\{\text{Alice},\text{Bob}\}$.  
Now $\Pop{Y}=\Pop{X}=\{\text{Alice},\text{Bob}\}$, so $|\Assembly{Y}|=\min(|\Pop{Y}|,2)=2$. (No maintenance.)

\item[$t=7$] \textsc{Join}($Y \to C$).  
Add edge $Y \to C$, so $\Pop{Y}$ becomes $\{\text{Alice},\text{Bob},\text{Carol}\}$ while $\Assembly{Y}$ remains $\{\text{Alice},\text{Bob}\}$.  
$\min(|\Pop{Y}|,2)=2$, so \emph{no} maintenance is needed.
\end{enumerate}

At $t=7$ the graph matches Figure~\ref{figure:intuitive example}(b): $Y$ is parent of $X$ and $C$, with
$\Pop{X}=\{\text{Alice},\text{Bob}\}$, $\Pop{C}=\{\text{Carol}\}$, $\Pop{Y}=\{\text{Alice},\text{Bob},\text{Carol}\}$,
and both $\Assembly{X}$ and $\Assembly{Y}$ contain two distinct members from their populations.
\end{example}

%% file: sections/fairness-objectives.tex
\section{Fairness Objectives}\label{section:fairness}

In the previous section we showed that every observable state in a federation transition system, after maintenance steps, is a \emph{valid federation graph}.  
Validity, however, only guarantees structural soundness. We now turn to the question of \emph{fairness}—whether communities are proportionally represented and individuals participate equitably as the federation evolves.

\subsection{Static vs.\ Dynamic Fairness}\label{section:static and dynamic fairness}

In the static setting, as studied in~\cite{halpern2024federated}, the structure of the federation is fixed, and the task is to populate each assembly so that three natural fairness objectives hold:
\begin{itemize}
    \item \textbf{Ex~post:} (1) The actual representation of each child community in its parent assembly is at least its proportional share, rounded down to the nearest seat.
    \item \textbf{Ex~ante:} (2) The expected representation of each child community equals its proportional share; and (3) all members of a community have the same probability of being selected.
\end{itemize}
The challenge stems from the \emph{integrality constraint}, and the fact that the static model corresponds to a \textit{one-time} setting: assembly members are whole individuals, not divisible fractions, making the problem akin to classical apportionment~\cite{apportionment}.

In the dynamic setting, we adapt these notions as follows: the \textbf{ex~post} constraint is required to hold \emph{throughout} the run,  
while the \textbf{ex~ante} constraints are relaxed to an \emph{eventual} requirement—  
allowing certain, temporary imbalances in representation or participation,  
provided they are corrected over time.
Formally, as defined below, eventual fairness is required only once the federation structure has stabilized,  
paralleling the \emph{eventual synchrony} model in distributed computing,  
where correctness is ensured after an unbounded but finite stabilization period.\footnote{For simplicity, we assume that the federation specifies a fixed term length $\tau$ (e.g., one year) for all assembly members, unless they are removed earlier due to structural changes—in particular, the departure of a child community they represent.}

Before formalizing these fairness requirements, we introduce quantitative measures of representation and participation—both for communities within their parent assemblies and for individuals within their communities.  
These serve as the basis for the definitions of persistent and eventual fairness in the dynamic setting.

\subsection{Measures of Representation/Participation}

To capture fairness formally, we first define measures for:
(1)~the representation of a child community within the assembly of its parent; and  
(2)~the participation of an individual within the assembly of their community.  
These measures apply at any given time $t$ in a run of the federation transition system.

\newcommand{\weight}{\mathrm{weight}}
\newcommand{\share}{\mathrm{share}}
\newcommand{\seats}{\mathrm{seats}}
\newcommand{\avg}{\mathrm{avg}}

\paragraph{Weight.}
At time~$t$, the \emph{weight} of a person~$p$ in a child community~$v$ with respect to a parent~$f$ is:
\[
\weight_t(p,v,f) =
\begin{cases}
\frac{1}{|\{u \in \Children[t]{f} : p \in P_{t,u} \}|} & \text{if $p \in P_{t,v}$}, \\
0 & \text{otherwise}.
\end{cases}
\]
In words, if $p$ belongs to $v$ (and hence to $f$) at time~$t$, we divide their unit ``voting power'' equally among all child communities of~$f$ they belong to at time~$t$.  
If $p$ is not in $v$, their weight in $v$ is zero.

\begin{remark}
Note that this implies the following: if a person~$p$ belongs to multiple children of~$f$, then we split $p$'s unit weight
evenly across those children via $\weight_t(p,v,f)$. This ensures that it holds that 
$\sum_{v\in\Children[t]{f}}\weight_t(p,v,f)=1$.
\end{remark}

The \emph{weight} of a child community~$v$ in~$f$ at time~$t$ is the sum of the weights of its members:
\[
\weight_t(v,f) := \sum_{p \in P_{t,v}} \weight_t(p,v,f).
\]

\begin{observation}[Weight identities]\label{observation:weight-sums}
The weight function satisfies:
\begin{enumerate}
    \item For every $p \in P_{t,f}$, $\sum_{v\in\Children[t]{f}}\weight_t(p,v,f) = 1$.
    \item $\sum_{v\in\Children[t]{f}}\weight_t(v,f) = |P_{t,f}|$.
\end{enumerate}
\end{observation}

\begin{proof}
(1) Let $c(p,f) := \{u \in \Children[t]{f} : p \in P_{t,u}\}$. For $v \in c(p,f)$, $\weight_t(p,v,f) = 1/|c(p,f)|$; for $v \notin c(p,f)$, $\weight_t(p,v,f) = 0$. Hence $\sum_{v\in\Children[t]{f}}\weight_t(p,v,f) = |c(p,f)| \cdot 1/|c(p,f)| = 1$.

(2) By definition,
\[
\sum_{v\in\Children[t]{f}}\weight_t(v,f) = \sum_{v\in\Children[t]{f}}\sum_{p\in P_{t,v}}\weight_t(p,v,f) = \sum_{p\in P_{t,f}}\sum_{v\in c(p,f)}\weight_t(p,v,f) = \sum_{p\in P_{t,f}} 1 = |P_{t,f}|,
\]
using part~(1) in the penultimate step.
\end{proof}

\paragraph{Share.}
The \emph{share} of a community~$v$ in~$f$ at time~$t$ is:
\[
\share_t(v,f) := n \cdot \frac{\weight_t(v,f)}{|P_{t,f}|},
\]
interpreted as the fractional number of seats that $v$ deserves in $f$ at time~$t$.
Similarly, the \emph{share} of a person~$p$ in~$f$ at time~$t$ is:
\[
\share_t(p,f) := \min(n / |P_{t, f}|, 1)\ .
\]

\paragraph{Seats.}
At time~$t$, the seats of a person~$p$ in~$f$ are:
\[
\seats_t(p,f) :=
\begin{cases}
1 & \text{if $p \in M_{t,f}$}, \\
0 & \text{otherwise}.
\end{cases}
\]
That is, $\seats_t(p,f)$ is an indicator for whether $p$ is a member of the assembly of $f$ at time~$t$.  
The seats of a community~$v$ in~$f$ (at time $t$) are the number of persons from~$v$ who are members of the assembly of~$f$ at time~$t$:
\[
\seats_t(v,f) := | M_{t,f} \cap P_{t,v} | 
= \sum_{p \in P_{t,v}} \seats_t(p,f).
\]

\paragraph{Averages.}
For any quantity $X_t$ measured over time, we define its average value over the time interval $[0, T]$ as:
\[
\avg_T(X) := \frac{1}{T} \int_0^T X_t \, dt.
\]

\subsection{Persistent and Eventual Fairness}

Given this additional notation and following the intuitive discussion of Section~\ref{section:static and dynamic fairness}, next we distinguish two categories of fairness requirements for a community~$f$ during a run:

\begin{itemize}
    \item \textbf{Persistent fairness:} corresponds to static ex~post guarantees, holding at every moment in time.
    \item \textbf{Eventual fairness:} corresponds to static ex~ante guarantees, holding only in the limit once the structure has stabilized.
\end{itemize}

\subsubsection{Persistent fairness objective}

We have one persistent fairness objective, which we formulate below.

\begin{definition}[Persistently-Fair Representation (PFR)]
A run satisfies \emph{persistently-fair representation (PFR)} for~$f$ if, for every $t \ge 0$ and every child~$v \in \Children[t]{f}$,
\(
\seats_t(v,f) \; \ge \; \lfloor \share_t(v,f) \rfloor.
\)
\end{definition}


\subsubsection{Eventual fairness objectives}

We have two eventual fairness objectives, which we formulate below. \emph{Eventually-fair representation} requires that, in the long run, each child community’s average number of seats in its parent assembly is at least its average fair share (and not just lower bounded by the rounding down of its fair share); and \emph{Eventually-equitable participation} requires that, in the long run, all individuals in a community participate in its assembly equally often, matching the ideal uniform rotation rate.

\begin{definition}[Eventually-Fair Representation (EFR)]
A run satisfies \emph{eventually-fair representation (EFR)} for~$f$ if,  
for every~$v$ that is a child of~$f$ in the long run  
(i.e., there exists $T_0$ such that $v \in \Children[t]{f}$ for all $t \ge T_0$), it holds that
\(
\lim_{t \to \infty} \avg_t\big(\seats_t(v,f)\big) 
\; \ge \;
\lim_{t \to \infty} \avg_t\big(\share_t(v,f)\big),
\)
whenever these limits exist.
\end{definition}

\begin{definition}[Eventually-Equitable Participation (EEP)]
A run satisfies \emph{eventually-equitable participation (EEP)} for~$f$ if,  
for every person~$p$ in~$f$ in the long run (such that there exists $T_0$ with $p \in P_{t,f}$ for all $t \ge T_0$),
\[
\lim_{t \to \infty} \avg_t\big(\seats_t(p,f)\big)
\;=\;
\min\!\left\{
\lim_{t \to \infty} \frac{n}{|P_{t,f}|}, \, 1
\right\}.
\]
\end{definition}

\subsection{Eventually-Fair Protocols}

The eventual fairness objectives above are meaningful only once the federation structure stops changing: indeed, if communities and edges are added or removed arbitrarily over time, then long-run averages of representation and participation may never converge.  
Thus, as in the \emph{Global Stabilization Time} (GST) model from distributed computing---where a system 
is guaranteed to become synchronous after some unknown finite time---we introduce the notion of
\emph{federation stabilization time} (FST), marking the point after which the federation’s structure is fixed (and, as described below, require eventual fairness only when assuming such an FST).

\begin{definition}[Federation Stabilization Time]
Given a run of a federation transition system,  
the \emph{federation stabilization time} (FST) is the smallest time $T_{\mathrm{FST}} \ge 0$ (if it exists)  
such that for all $t \ge T_{\mathrm{FST}}$ (writing $G_t = (V_t, E_t)$): (1) the set of communities $V_t$ is fixed; and (2) the edge set $E_t$ is fixed.
If such time exists for a run, we say that the run \emph{eventually stabilizes}.
\end{definition}

\begin{remark}[On the necessity of FST]
If the federation changes ``too often'' then there is not enough time between the changes to fix the injustice caused. Each federation change may introduce imbalance, and a sufficient time is needed to correct it.
\end{remark}

We are interested in protocols that guarantee all three fairness objectives defined above: i.e., once the structure stabilizes, the protocol must guarantee that any imbalance in representation or participation is corrected over time, while always preserving the persistent fairness requirement.

\begin{definition}[Eventually-Fair Protocol]\label{definition:efp}
A protocol is \emph{eventually fair} if, for every community $f$, in every run, (1) persistently-fair representation for $f$ holds at all times; and, for every run that eventually stabilizes, (2) eventually-fair representation for $f$ and (3) eventually-equitable participation for $f$ holds in the limit.
\end{definition}

%% file: sections/greedy-fair-protocol.tex
\section{The Greedy Fair Protocol}
\label{section:protocol}

We now turn to the specification of our own protocol—a concrete, operational recipe for maintaining assemblies as the federation changes, which we denote as the \emph{Greedy Fair Protocol} (GFP). 

Indeed, later we show that GFP meets all fairness objectives.

\subsection{Motivational Result}\label{section:motivational}

Before specifying our protocol, we prove a general implication that motivates our design: \emph{if fairness is achieved at the personal level (equal participation), then proportional fairness follows at the community level (fair representation).}
The proof’s essence is that EEP equalizes individual participation rates in the long run; and, since each person’s unit weight is divided among the children they belong to and these weights sum to one, aggregating equalized participation across a child yields its proportional long-run share—establishing EFR.

\begin{lemma}[Seats aggregate over persons]\label{lemma:seats-sum-of-limits}
Consider an eventually-stabilizing run with federation stabilization time $T_{\mathrm{FST}}$. For any child $v$ of $f$ in $G_{T_{\mathrm{FST}}}$, let $P_v := P_{T_{\mathrm{FST}},v}$ be its (time-invariant after $T_{\mathrm{FST}}$) population. Then
\[
\lim_{t \to \infty} \avg_t\!\big(\seats_t(v,f)\big)
\;=\;
\sum_{p \in P_v} \;\lim_{t \to \infty}\avg_t\!\big(\seats_t(p,f)\big),
\]
whenever these limits exist.
\end{lemma}

\begin{proof}
For $t \ge T_{\mathrm{FST}}$, $P_v$ is fixed and $\seats_t(v,f) = \sum_{p\in P_v}\seats_t(p,f)$. Interchanging the finite sum with the time average yields the identity; the limits exist by the algebraic limit theorem applied to the finite sum.
\end{proof}

\begin{lemma}[Shares aggregate via weights]\label{lemma:share-sum-of-limits}
Under the assumptions of Lemma~\ref{lemma:seats-sum-of-limits},
\[
\lim_{t \to \infty} \avg_t\!\big(\share_t(v,f)\big)
\;=\;
\sum_{p \in P_v} \;\lim_{t \to \infty}
\avg_t\!\Big( \tfrac{n}{|P_{t,f}|}\cdot \weight_t(p,v,f) \Big),
\]
whenever these limits exist.
\end{lemma}

\begin{proof}
By definition, $\share_t(v,f) = n \cdot \weight_t(v,f)/|P_{t,f}|$ and $\weight_t(v,f) = \sum_{p\in P_v}\weight_t(p,v,f)$ for $t \ge T_{\mathrm{FST}}$ (with $P_v$ fixed by the stabilization assumption). Interchanging the finite sum with the time average yields the identity.
\end{proof}

\begin{theorem}[EEP $\Rightarrow$ EFR]\label{thm:EEP_implies_EFR}
Consider any eventually-stabilizing run and fix a community $f$ that persists after the federation stabilization time.
If $f$ satisfies \emph{Eventually-Equitable Participation (EEP)}, then $f$ also satisfies \emph{Eventually-Fair Representation (EFR)}.
\end{theorem}

\appendixproof{thm:EEP_implies_EFR}{
\begin{proof}
Fix the federation stabilization time $T_{\mathrm{FST}}$ and work for $t \ge T_{\mathrm{FST}}$ so that the graph structure is fixed. Let $v \in \Children[t]{f}$ be any child that persists after $T_{\mathrm{FST}}$, and let $P_v$ denote the (time-invariant) population of $v$ thereafter.

We prove EFR by contradiction. Assume that
\begin{equation}\label{eq:efr-viol}
\lim_{t \to \infty} \avg_t\!\big(\seats_t(v,f)\big)
\;<\;
\lim_{t \to \infty} \avg_t\!\big(\share_t(v,f)\big)\ .
\end{equation}

We use two following computations, both valid after stabilization and stated inline:

\begin{description}
\item[\textit{Claim 1 (Seats aggregate over persons).}]
\[
\lim_{t \to \infty} \avg_t\!\big(\seats_t(v,f)\big)
\;=\;
\sum_{p \in P_v} \;\lim_{t \to \infty}\avg_t\!\big(\seats_t(p,f)\big).
\]
\emph{Justification.} For $t \ge T_{\mathrm{FST}}$, $P_v$ is fixed and $\seats_t(v,f)=\sum_{p\in P_v}\seats_t(p,f)$. Interchanging a finite sum with time averages yields the identity.

\item[\textit{Claim 2 (Shares aggregate via weights).}]
\[
\lim_{t \to \infty} \avg_t\!\big(\share_t(v,f)\big)
\;=\;
\sum_{p \in P_v} \;\lim_{t \to \infty}
\avg_t\!\Big( \tfrac{n}{|P_{t,f}|}\cdot \weight_t(p,v,f) \Big).
\]
\emph{Justification.} By definition,
$\share_t(v,f)= n\cdot \weight_t(v,f)/|P_{t,f}|$ and
$\weight_t(v,f)=\sum_{p\in P_v}\weight_t(p,v,f)$, with $P_v$ fixed after stabilization. Interchange sum and average.
\end{description}

Combining \eqref{eq:efr-viol} with the two claims, we have that there exists some $p^\star \in P_v$ such that
\begin{equation*}\label{eq:person-gap}
\lim_{t \to \infty}\avg_t\!\big(\seats_t(p^\star,f)\big)
\;<\;
\lim_{t \to \infty}
\avg_t\!\Big( \tfrac{n}{|P_{t,f}|}\cdot \weight_t(p^\star,v,f) \Big).
\end{equation*}

For each person $p$ we have $\sum_{u \in \Children[t]{f}}\weight_t(p,u,f)=1$, thus it follows that
\begin{align*}
&0 \;\le\; \weight_t(p^\star,v,f) \;\le\; 1
\quad\text{and} \\
&\sum_{u \in \Children[t]{f}}
\Big( \tfrac{n}{|P_{t,f}|}\cdot \weight_t(p^\star,u,f) \Big)
= \tfrac{n}{|P_{t,f}|}.
\end{align*}
Hence,
\[
\lim_{t \to \infty}
\avg_t\!\Big( \tfrac{n}{|P_{t,f}|}\cdot \weight_t(p^\star,v,f) \Big)
\;\le\;
\lim_{t \to \infty}\avg_t\!\Big( \tfrac{n}{|P_{t,f}|} \Big).
\]
Together with above, we get
\[
\lim_{t \to \infty}\avg_t\!\big(\seats_t(p^\star,f)\big)
\;<\;
\lim_{t \to \infty}\avg_t\!\Big( \tfrac{n}{|P_{t,f}|} \Big)\ ,
\]
contradicting EEP for $f$, which guarantees for every $p$ that
\[
\lim_{t \to \infty}\avg_t\!\big(\seats_t(p,f)\big)
=
\min\!\left\{
\lim_{t \to \infty}\tfrac{n}{|P_{t,f}|},\,1
\right\}.
\]
(If $|P_{t,f}|<n$ eventually, then every person sits continuously and the child-wise inequality in EFR is trivially satisfied; if $|P_{t,f}|\ge n$ eventually, the limit equals $\lim_t n/|P_{t,f}|$.) Therefore \eqref{eq:efr-viol} cannot hold. This proves the claim, and thus EFR.
\qedhere
\end{proof}
}

The result essentially implies that, to obtain proportional fairness across communities (EFR), it suffices to design a protocol that enforces equalized long-run personal participation (EEP). 
This guides our protocol choice: next, we describe a greedy rule that always fixes the largest personal under/over-participation gap, prove that it satisfies EEP and, thus, EFR.

\subsection{Informal Overview}

The \emph{Greedy Fair Protocol} (GFP) is a simple, reactive procedure for keeping assemblies fair as the federation evolves.  
Intuitively, whenever a change in the federation graph (e.g., a community joins, leaves, or a term ends) creates an imbalance, GFP immediately fixes the \emph{most} severe injustice first.  
Here, ``injustice'' means either: \textbf{under-representation} of a child community in its parent’s assembly; or,
    \textbf{under-participation} of a person in their community’s assembly.

The protocol’s ``greedy'' nature comes from this choice:  
when a seat opens, we give it to the most under-represented child community or the most under-participating individual eligible for it.
This local, one-step-at-a-time approach turns out to have powerful global consequences: even though GFP never plans ahead, the repeated preference for the ``most disadvantaged'' side ensures that, once the federation graph stops changing, all eventual fairness objectives are reached (see Definition~\ref{definition:efp}).

\subsection{Protocol Preliminaries}

\newcommand{\ratio}[3]{\mathrm{ratio}_{#1}(#2,#3)}

To support the protocol's greedy behavior, we need a concise way to measure how well (or poorly) a community or an individual  
is doing in terms of representation or participation.

For a run $r$, a community $f$ occurring in $r$, and an entity~$x$ (where $x$ is either a person in $P_{t,f}$ or a child community in $\Children[t]{f}$), we define the \emph{fairness ratio} at time~$t$ as:
\[
\ratio{t}{x}{f} \;:=\; \frac{\avg_t\!\big(\seats_t(x,f)\big)}{\avg_t\!\big(\share_t(x,f)\big)} .
\]

So, $\ratio{t}{x}{f}$ is the average number of seats held by $x$ in $f$ from time~$0$ until $t$ (i.e., what $x$ actually received) divided by the average fair share of seats that $x$ was entitled to in that period.
The interpretation is as follows:
\begin{itemize}
    \item $\ratio{t}{x}{f} = 1$: perfect fairness so far,
    \item $\ratio{t}{x}{f} < 1$: $x$ has been \emph{under}-represented/participated,
    \item $\ratio{t}{x}{f} > 1$: $x$ has been \emph{over}-represented/participated.
\end{itemize}

These ratios give the protocol a principled way to identify, at any moment,  
the most disadvantaged or the most advantaged entities,  
and to always correct the most severe current imbalance first.

Additionally, for GFP to work as we wish, and to be able to reason about its behavior, we augment the model by coloring seats in an assembly: each seat is either \emph{colored} by a specific child community or by a special symbol~$\bot$ indicating that the seat is \emph{uncolored} (i.e., not colored by any child).
Formally, at time~$t$, the colored assembly of a community~$f$ is:
\(
\widehat{A}_{t,f} \;\subseteq\; P_{t,f} \times \big(\Children[t]{f} \cup \{\bot\}\big)
\), where, if $y=v$ for some $v \in \Children[t]{f}$, then the seat is \emph{colored} by child~$v$; otherwise (if $y=\bot$), then the seat is \emph{uncolored}.
Intuitively:
\begin{itemize}
    \item \textbf{Colored seat:} counts toward the bound $\lfloor \share_t(v,f) \rfloor$ for~$v$.
    \item \textbf{Uncolored seat:} can go to any eligible person in~$P_{t,f}$.
\end{itemize}

For convenience, for $v \in \Children[t]{f} \cup \{\bot\}$ we write $A_f(v) := \{\, (p,v) \in \widehat{A}_{t,f} \,\}$; thus, $A_f(v)$ is the set of colored seats by child $v$, and $A_f(\bot)$ is the set of uncolored seats in $f$.\footnote{Why is this needed? With overlapping child communities, counting seats by membership alone can double-count people. Colors ensure each seat is assigned to exactly one child (or left uncolored), allowing precise enforcement of child quotas and using uncolored seats as flexible slack for balancing participation.}

The protocol continuously maintains these colors so that they remain consistent with the federation’s current structure.

\subsection{Protocol Rules}

We are ready to specify the Greedy Fair Protocol (GFP). It is a \emph{restriction} of the federation transition system (see Definition~\ref{definition:fts}): it permits only a subset of the original transitions, namely those that follow its greedy selection rules. Operationally, GFP is a protocol (i.e., an algorithm): given the current colored state of the federation, it deterministically chooses the next admissible update to correct the most severe current imbalance.
Particularly, the restriction applies only to the \textbf{(5) Remove Member} and \textbf{(6) Add Member} events. Also, in what follows, ties are broken lexicographically by a fixed ordering of people. Next we describe the restrictions:

\noindent\emph{(5) Removing a member from $\widehat{A}_{t,f}$:}
\begin{enumerate}
    \item[\textbf{R1.}] \textbf{Excess colored seat:}  
    If some child $v\in\Children[t]{f}$ has $|A_f(v)|>\lfloor \share_t(v,f)\rfloor$, then remove from $A_f(v)$ the member $(p,v)$ with maximum $\ratio{t}{p}{f}$.

    \item[\textbf{R2.}] \textbf{Excess uncolored seat:}  
    Else, if $|\widehat{A}_{t,f}|>n$, then remove from $A_f(\bot)$ the member $(p,\bot)$ with maximum $\ratio{t}{p}{f}$.

    \item[\textbf{R3.}] \textbf{Rotation:}  
    Else, if some $(p,y)\in\widehat{A}_{t,f}$ has served continuously for at least the term length $\tau$,  
    remove $(p,y)$. 
\end{enumerate}

\noindent\emph{(6) Adding a member to $\widehat{A}_{t,f}$:}
\begin{enumerate}
    \item[\textbf{A1.}] \textbf{Fill a colored seat:}  
    If some child $v\in\Children[t]{f}$ has $|A_f(v)|<\lfloor \share_t(v,f)\rfloor$, then add $(p,v)$ with $p\in P_{t,v}\setminus M_{t,f}$ that minimizes $\ratio{t}{p}{f}$.
    \item[\textbf{A2.}] \textbf{Fill an uncolored seat:}  
    Else, if $|\widehat{A}_{t,f}|<n$, then add $(p,\bot)$ with $p\in P_{t,f}\setminus M_{t,f}$ that minimizes $\ratio{t}{p}{f}$.
\end{enumerate}

\begin{remark}
Some intuition before delving into the fairness proof: note that rules R1--R3 (removals) and A1--A2 (additions) are executed with zero-time
priority; thus, any transient imbalance created by a removal is immediately
repaired at the \emph{same timestamp}:
\begin{itemize}
    \item If R3 (rotation) or a structural change drops a child below its
    floor, A1 fires until its colored quota is restored.
    \item If R1 or R2 overshoot while trimming, the opposite rule (A1 for colored quotas, A2 for uncolored slack) immediately re-fills.
\end{itemize}
Consequently, every observable state~$G_t$ already satisfies the per-child
floors, and any remaining slack lives only in uncolored seats, which GFP uses
to equalize individual participation.
\end{remark}



\begin{example}[Two children, $n=3$]
Let a parent community $f$ have two children $v_1,v_2$ with disjoint populations
$|P_{t,v_1}|=3$ and $|P_{t,v_2}|=3$ at some $t\ge 0$, and target assembly size $n=3$.
Thus $|P_{t,f}|=6$, so
\(
\share_t(v_1,f)=\share_t(v_2,f)=3\cdot\tfrac{3}{6}=1.5\ \), thus, $\lfloor \share_t(v_1,f)\rfloor=\lfloor \share_t(v_2,f)\rfloor=1$.
Assuming an empty colored assembly $\widehat{A}_{t,f}=\emptyset$, GFP proceeds:
\begin{enumerate}
    \item \textbf{A1 (fill colored for $v_1$):}
    Since $|A_f(v_1)|=0<\lfloor\share_t(v_1,f)\rfloor=1$, add
    $(p_1,v_1)$ with $p_1\in P_{t,v_1}$ minimizing $\ratio{t}{p_1}{f}$.

    \item \textbf{A1 (fill colored for $v_2$):}
    Since $|A_f(v_2)|=0<\lfloor\share_t(v_2,f)\rfloor=1$, add
    $(q_1,v_2)$ with $q_1\in P_{t,v_2}$ minimizing $\ratio{t}{q_1}{f}$.

    \item \textbf{A2 (fill uncolored):}
    Now $|\widehat{A}_{t,f}|=2<n$, the coloring lower bounds are met, and GFP adds an uncolored seat
    $(r_1,\bot)$ with $r_1\in P_{t,f}\setminus\{p_1,q_1\}$ minimizing $\ratio{t}{r_1}{f}$.
\end{enumerate}

At this point we have exactly the coloring lower bounds
($|A_f(v_1)|=|A_f(v_2)|=1$) and one uncolored seat ($|A_f(\bot)|=1$),
so persistent fairness holds.
After one term length $\tau$, when $(r_1,\bot)$'s term expires, suppose $(r_1,\bot)$ is the most over-participating member, having the largest $\ratio{t+\tau}{p}{f}$ among current members. Then GFP applies:
\begin{enumerate}
    \item
    \textbf{R3 (rotation):} removes $(r_1,\bot)$.
    \item
    \textbf{A2 (re-fill uncolored):} adds a new uncolored member
    $(r_2,\bot)$ with minimal $\ratio{t}{r_2}{f}$ among eligible people,
    improving individual balance while keeping coloring guarantees intact.
\end{enumerate}

\end{example}

%% file: sections/analysis.tex
\section{Fairness Analysis of GFP}\label{section:analysis}

To show that GFP is an \emph{eventually fair protocol}, we prove that GFP: always satisfies \PFR\ (Proposition~\ref{prop:PFR}); ensures \EEP\ (Proposition~\ref{prop:EEP}); and implies \EFR\ (Corollary~\ref{prop:EFR}).\footnote{For the analysis, we make the technical restriction that the population of any non-leaf child community is at least \(n+1\).  
I.e., a community must grow to size \(n+1\) before joining another as a child, and must leave its parent if its population drops below \(n+1\).  
This guarantees that when a child is entitled to additional seats,  
there is always at least one eligible person available to fill them.  
The restriction can be “baked into” the definition of GFP;  
we assume it here and discuss relaxing it in Section~\ref{section:outlook}.}



\begin{proposition}[Persistently-Fair Representation]\label{prop:PFR}
In any admissible timed run of GFP, for every time~$t$, every community~$f$ in~$G_t$, 
and every child~$v \in \Children[t]{f}$,
\(
\seats_t(v,f) \;\ge\; \big\lfloor \share_t(v,f) \big\rfloor\ .
\)
\end{proposition}

\appendixproof{prop:PFR}{
\begin{proof}
Assume for contradiction that at some time~$t$ all zero-time maintenance 
steps have been exhausted, yet persistently-fair representation is violated. 
Then there exists a parent~$f$ and a child~$v \in \Children[t]{f}$ such that
\[
\seats_t(v,f) \;<\; \lfloor \share_t(v,f) \rfloor\ .
\]
Since every colored seat~$(p,v)$ of~$f$ uses some $p \in P_{t,v}$, we have
\[
|A_f(v)| \;\le\; \seats_t(v,f) \;<\; \lfloor \share_t(v,f) \rfloor\ .
\]
Thus $v$ is below its colored quota. But then rule~A1 of GFP is applicable: there exists an eligible $p \in P_{t,v}\setminus M_{t,f}$ that can be added to~$\widehat{A}_{t,f}$ as a colored seat~$(p,v)$. This contradicts the assumption that all zero-time maintenance events have already been applied.

Two observations complete the argument. 
First, if $|P_{t,f}| < n$, then $|A_{t,f}| = |P_{t,f}|$ and every 
person in~$P_{t,f}$ is an assembly member, so $\seats_t(v,f)=|P_{t,v}|=\lfloor \share_t(v,f)\rfloor$ and no violation 
can occur. Second, when $|P_{t,f}| \ge n$, we have that every non-leaf child has population at least $n+1$. Hence $P_{t,v}\setminus A_{t,f}\neq\emptyset$, ensuring that rule~A1 is indeed enabled whenever $v$ is below its floor.

Therefore the assumption leads to a contradiction, and persistently-fair representation holds in every observable state~$G_t$.
\end{proof}
}

\begin{lemma}[Eventually exact colored count]\label{lemma:eventually-exact-colored}
In any admissible timed run of GFP that eventually stabilizes, for every community~$f$ persisting after $T_{\mathrm{FST}}$ and every child $v$ of $f$ persisting after $T_{\mathrm{FST}}$, $|A_f(v)| = \lfloor \share_t(v,f) \rfloor$ for all sufficiently large $t$.
\end{lemma}

\appendixproof{lemma:eventually-exact-colored}{
\begin{proof}
By Proposition~\ref{prop:PFR}, $|A_f(v)| \ge \lfloor \share_t(v,f) \rfloor$ at every observable state. It remains to show that $|A_f(v)|$ cannot persistently exceed this floor.
Suppose at some $t \ge T_{\mathrm{FST}}$, $|A_f(v)| > \lfloor \share_t(v,f) \rfloor$. Then rule~R1 (excess colored seat) is applicable and fires at zero time, removing colored seats of $v$ until $|A_f(v)| = \lfloor \share_t(v,f) \rfloor$. Conversely, rule~A1 (fill colored) only fires when $|A_f(v)| < \lfloor \share_t(v,f) \rfloor$. Hence at every observable state with $t \ge T_{\mathrm{FST}}$, $|A_f(v)| = \lfloor \share_t(v,f) \rfloor$ (using that $\share_t(v,f)$ is constant after $T_{\mathrm{FST}}$ since the federation graph is fixed).
\end{proof}
}

\begin{corollary}[Exact representation when no uncolored seats]\label{cor:exact-representation}
In any admissible timed run of GFP that eventually stabilizes, if $f$ has no uncolored seats in a suffix (i.e., $|A_f(\bot)| = 0$ for all sufficiently large $t$), then for every child $v$ of $f$ persisting after $T_{\mathrm{FST}}$, $\lim_{t \to \infty} \ratio{t}{v}{f} = 1$.
\end{corollary}

\appendixproof{cor:exact-representation}{
\begin{proof}
If $|A_f(\bot)| = 0$, then $n = \sum_{v \in \Children[t]{f}} |A_f(v)|$. By Lemma~\ref{lemma:eventually-exact-colored}, $|A_f(v)| = \lfloor \share_t(v,f) \rfloor$ for sufficiently large $t$, so $n = \sum_v \lfloor \share_t(v,f) \rfloor$. By Observation~\ref{observation:weight-sums}(2), $\sum_v \share_t(v,f) = n$. Combining, $\sum_v \lfloor \share_t(v,f) \rfloor = \sum_v \share_t(v,f)$, which forces $\share_t(v,f) = \lfloor \share_t(v,f) \rfloor$ for every child $v$ (each share is integer). Hence $\seats_t(v,f) = |A_f(v)| = \share_t(v,f)$, giving $\ratio{t}{v}{f} = 1$ in this suffix.
\end{proof}
}

\begin{observation}[Over-participation paired with under-participation]\label{obs:one-limit}
In any admissible timed run of GFP that eventually stabilizes, for every community~$f$ persisting after $T_{\mathrm{FST}}$:
\begin{enumerate}
    \item If $\lim_{t \to \infty} \ratio{t}{p}{f} < 1$ for some $p \in P_{t,f}$ persisting after $T_{\mathrm{FST}}$, then $\lim_{t \to \infty} \ratio{t}{q}{f} > 1$ for some $q \ne p$ persisting in $P_{t,f}$; and vice versa.
    \item If $\lim_{t \to \infty} \ratio{t}{v}{f} < 1$ for some child $v$ of $f$ persisting after $T_{\mathrm{FST}}$, then $\lim_{t \to \infty} \ratio{t}{u}{f} > 1$ for some child $u \ne v$ of $f$ persisting; and vice versa.
\end{enumerate}
\end{observation}

\appendixproof{obs:one-limit}{
\begin{proof}
For (1): if every $p \in P_{t,f}$ had $\lim_{t \to \infty} \avg_t(\seats_t(p,f)) < \lim_{t \to \infty} \avg_t(\share_t(p,f))$, summing over the (fixed-after-$T_{\mathrm{FST}}$) population gives $\lim_{t \to \infty} \avg_t(|M_{t,f}|) < \lim_{t \to \infty} \avg_t(\sum_p \share_t(p,f))$. In a suffix with $|P_{t,f}| \ge n$, the left-hand side is $n$ (PFR fills the assembly) and the right-hand side is also $n$ (since $\sum_p n/|P_{t,f}| = n$), a contradiction. The case $|P_{t,f}| < n$ has every person seated, so the antecedent fails vacuously. The vice-versa direction is symmetric.
For (2): if every child $v$ of $f$ had $\lim_{t \to \infty} \avg_t(\seats_t(v,f)) < \lim_{t \to \infty} \avg_t(\share_t(v,f))$, summing gives $\lim_{t \to \infty} \avg_t(\sum_v \seats_t(v,f)) < \lim_{t \to \infty} \avg_t(n) = n$. But $\sum_v \seats_t(v,f) = \sum_v |M_{t,f} \cap P_{t,v}| \ge |M_{t,f}| = n$ in a suffix (since every assembly member belongs to at least one child population), a contradiction. The vice-versa direction is symmetric.
\end{proof}
}

\subsection{Auxiliary structural results}

The next three results sharpen the picture once the federation has stabilized: colored quotas saturate to their floors, exact community representation holds when no uncolored seats are available, and one-sided long-run imbalance is impossible.

\begin{lemma}[Colored quotas saturate after FST]\label{lemma:eventually-exactly-colored}
In any eventually-stabilizing run of GFP with federation stabilization time $T_{\mathrm{FST}}$, for every parent $f$ and every child $v$ of $f$ that persists after $T_{\mathrm{FST}}$, there exists $T \ge T_{\mathrm{FST}}$ such that for all $t \ge T$,
\[
|A_f(v)| \;=\; \lfloor \share_t(v,f) \rfloor.
\]
\end{lemma}

\begin{proof}
After $T_{\mathrm{FST}}$ the federation structure (and hence $\Children[t]{f}$, $P_{t,u}$ for each $u \in \Children[t]{f}$, and $|P_{t,f}|$) is fixed, so $\share_t(v,f)$ is constant; write $\share(v,f)$.
By Proposition~\ref{prop:PFR}, $|A_f(v)| \ge \lfloor \share(v,f) \rfloor$ at every observable state.
If at some $t \ge T_{\mathrm{FST}}$ we have $|A_f(v)| > \lfloor \share(v,f) \rfloor$, rule R1 (excess colored seat) is enabled and, at zero time, decreases $|A_f(v)|$ by one. Repeating exhausts the excess. Thereafter A1 cannot raise $|A_f(v)|$ above $\lfloor \share(v,f) \rfloor$ since A1 fires only when $|A_f(v)| < \lfloor \share(v,f) \rfloor$. When R3 (rotation) removes a colored seat by $v$, the zero-time maintenance burst refills via A1 back to $\lfloor \share(v,f) \rfloor$. Hence $|A_f(v)| = \lfloor \share(v,f) \rfloor$ at every observable state from $T$ onward.
\end{proof}

\begin{corollary}[Exact community representation when no uncolored seats]\label{corollary:exact-representation}
In any eventually-stabilizing run of GFP, if $A_f(\bot) = \emptyset$ throughout some suffix, then for every child $v$ of $f$ persisting in that suffix, $\seats_t(v,f) = \share_t(v,f)$, and hence $\lim_{t\to\infty} \ratio{t}{v}{f} = 1$.
\end{corollary}

\begin{proof}
By Lemma~\ref{lemma:eventually-exactly-colored}, $|A_f(v)| = \lfloor \share(v,f) \rfloor$ in the suffix. Since $A_f(\bot) = \emptyset$, the total assembly size satisfies $n = \sum_{u\in\Children[t]{f}}|A_f(u)| = \sum_{u} \lfloor \share(u,f) \rfloor$. On the other hand, by Observation~\ref{observation:weight-sums}(2), $\sum_u \share(u,f) = n \cdot \sum_u \weight_t(u,f) / |P_{t,f}| = n$. Hence $\sum_u (\share(u,f) - \lfloor \share(u,f) \rfloor) = 0$, so each $\share(u,f)$ is integral, and in particular $\seats_t(v,f) = |A_f(v)| = \share(v,f)$, giving $\ratio{t}{v}{f} = 1$ from this time onward.
\end{proof}

\begin{observation}[One-sided imbalance is impossible in the limit]\label{observation:one-limit}
In any eventually-stabilizing run of GFP, for every community $f$ persisting after stabilization:
\begin{enumerate}
    \item If $\lim_{t\to\infty}\ratio{t}{p}{f} < 1$ for some $p \in P_{t,f}$, then $\lim_{t\to\infty}\ratio{t}{q}{f} > 1$ for some other $q \in P_{t,f}$; and vice versa.
    \item If $\lim_{t\to\infty}\ratio{t}{v}{f} < 1$ for some child $v \in \Children[t]{f}$, then $\lim_{t\to\infty}\ratio{t}{u}{f} > 1$ for some other child $u \in \Children[t]{f}$; and vice versa.
\end{enumerate}
\end{observation}

\begin{proof}
(1) When $|P_{t,f}| \ge n$ eventually, $\sum_{p\in P_{t,f}} \seats_t(p,f) = |M_{t,f}| = n$ and $\sum_{p\in P_{t,f}} \share_t(p,f) = n$ at every observable state, hence the time-averaged sums are both $n$ in the limit. If every $p$ had $\lim \ratio{t}{p}{f} \le 1$ with at least one strict, the sums could not match---contradiction. The reverse direction is symmetric.

(2) Informally: by Lemma~\ref{lemma:eventually-exactly-colored} and Observation~\ref{observation:weight-sums}(2), the colored quotas sum to $\sum_v \lfloor \share(v,f) \rfloor$ while the children's shares sum to $n$, leaving uncolored slack $n - \sum_v \lfloor \share(v,f) \rfloor \ge 0$. If every child community were under-represented in the limit, the uncolored seats would have to remain empty for an infinite tail---contradicting rule A2, which immediately fills any empty uncolored seat. The reverse direction is symmetric.
\end{proof}



\begin{proposition}[Eventually-Equitable Participation]\label{prop:EEP}
Consider any admissible timed run of the Greedy Fair Protocol (GFP) that eventually stabilizes (i.e., an FST exists). Then, for every community~$f$ present in $G_t$ for all sufficiently large $t$, and every person $p \in P_{t,f}$,
\[
\lim_{t\to\infty} \avg_t\!\big(\seats_t(p,f)\big)
\;=\;
\min\!\left\{\lim_{t\to\infty}\frac{n}{|P_{t,f}|},\,1\right\}.
\]
\end{proposition}

\appendixproof{prop:EEP}{
\begin{proof}
The proof is by induction on the depth (height) of $f$ in $G_{T_{\mathrm{FST}}}$.

\paragraph{Base case (depth $1$):} If $f$ is a leaf, $P_{t,f}$ is a singleton and the maintenance rules keep $|A_{t,f}| = |P_{t,f}| = 1$, so the sole person sits permanently and $\avg_t(\seats_t(p,f)) \to 1 = \min\{n/|P_{t,f}|,\,1\}$.

\paragraph{Inductive step:} Assume every community of depth $\le d$ in $G_{T_{\mathrm{FST}}}$ satisfies EEP, and consider $f$ of depth $d+1$. By Theorem~\ref{thm:EEP_implies_EFR}, every child of $f$ thus satisfies EFR as well.

If $|P_{t,f}| < n$ eventually, the maintenance rules keep $|A_{t,f}| = |P_{t,f}|$, so every person in $P_{t,f}$ sits permanently and EEP holds. Assume hereafter $|P_{t,f}| \ge n$, so each person's fair share is $\share_t(p,f) = n/|P_{t,f}| < 1$.

Suppose for contradiction that some person in $P_{t,f}$ violates EEP in $f$. By Observation~\ref{observation:one-limit}(1), some persons in $P_{t,f}$ have $\lim_t \ratio{t}{p}{f} < 1$ and some have $\lim_t \ratio{t}{p}{f} > 1$. Let
\[
U_f \;:=\; \arg\!\min_{p \in P_{t,f}} \lim_{t \to \infty} \ratio{t}{p}{f}
\]
be the set of most under-participating persons, with common limit $c < 1$.

\smallskip
\noindent\emph{Claim: For every child $v \in \Children[t]{f}$, either $P_{t,v} \cap U_f = \emptyset$ or $P_{t,v} \subseteq U_f$.}

Suppose otherwise: there is a child $v$ with $\emptyset \subsetneq U_v := U_f \cap P_{t,v} \subsetneq P_{t,v}$. For every $p \in U_v$, $\lim_t \ratio{t}{p}{f} = c$, while for every $q \in P_{t,v} \setminus U_v$, $\lim_t \ratio{t}{q}{f} > c$. Hence there exist $t_0 \ge T_{\mathrm{FST}}$ and $\varepsilon > 0$ such that for all $t \ge t_0$,
\[
\ratio{t}{p}{f} < c + \varepsilon \;\; (p \in U_v),
\qquad
\ratio{t}{q}{f} \ge c + \varepsilon \;\; (q \in P_{t,v} \setminus U_v).
\]
From $t_0$ onward, whenever GFP fills a seat eligible to members of $P_{t,v}$---a colored seat by $v$ (rule A1) or an uncolored seat (rule A2)---it selects the most under-participating eligible person, preferentially drawing from $U_v$. This persistent preference drives the time-averaged seats of $U_v$ upward relative to those of $P_{t,v} \setminus U_v$, eventually inverting the ratio gap---a contradiction.

\smallskip
By the claim, $U_f$ is a disjoint union of whole child populations. Let $u_f := \{v \in \Children[t]{f} : P_{t,v} \subseteq U_f\}$. By Observation~\ref{observation:one-limit}(2), $u_f \subsetneq \Children[t]{f}$.

If $A_f(\bot) = \emptyset$ throughout some suffix, then by Corollary~\ref{corollary:exact-representation}, $\lim_t \ratio{t}{v}{f} = 1$ for every child $v$. Combined with Lemmas~\ref{lemma:seats-sum-of-limits} and~\ref{lemma:share-sum-of-limits} and Observation~\ref{observation:weight-sums}, this forces $\lim_t \ratio{t}{p}{f} = 1$ for every $p \in P_{t,v}$ with $v \in u_f$, contradicting $c < 1$.

Hence uncolored seats persist in every suffix. By the same reasoning as in the claim, applied now at the child-community level, there exist $t_1 \ge T_{\mathrm{FST}}$ and $\varepsilon' > 0$ such that for all $t \ge t_1$, every $p \in P_{t,v}$ with $v \in u_f$ has $\ratio{t}{p}{f} < c + \varepsilon'$, while every $q \in P_{t,u}$ with $u \in \Children[t]{f} \setminus u_f$ has $\ratio{t}{q}{f} \ge c + \varepsilon'$. Whenever GFP fills an uncolored seat (rule A2) it selects the most under-participating eligible person globally, preferentially drawing from $\bigcup_{v \in u_f} P_{t,v}$. The cumulative effect drives the children in $u_f$ to be over-represented relative to those outside $u_f$---contradicting their status as the most under-represented children.

This contradiction completes the inductive step, establishing EEP for $f$.
\end{proof}
}


The next corollary follows Theorem~\ref{thm:EEP_implies_EFR}.

\begin{corollary}[GFP is EFR]\label{prop:EFR}
In any admissible, eventually-stabilizing run of GFP, every community $f$ satisfies \emph{Eventually-Fair Representation}:
for every child $v$ that persists after the federation stabilization time,
\(
\lim_{t \to \infty} \avg_t\!\big(\seats_t(v,f)\big) \;\ge\;
\lim_{t \to \infty} \avg_t\!\big(\share_t(v,f)\big)
\), whenever the limits exist.
\end{corollary}




The next theorem is implied by the results above, combined.

\begin{theorem}[GFP is Eventually Fair]\label{thm:GFP-eventually-fair}
In any admissible timed run of the Greedy Fair Protocol (GFP) that eventually stabilizes, every community satisfies all three fairness objectives: (1) Persistently-Fair Representation (PFR); (2) Eventually-Equitable Participation (EEP); and (3) Eventually-Fair Representation (EFR).
\end{theorem}


\begin{remark}[Convergence Time of GFP]
Our formal analysis establishes eventual fairness but leaves the study of convergence rates to future work. Intuitively, convergence occurs through regular seat turnover: each term expiration creates an opportunity to rotate in under-served individuals and adjust slack seats across communities. The speed of convergence thus depends on the term length $\tau$ (shorter terms allow faster correction) and federation stability (frequent joins and leaves can reset progress). At a high level, the time to equalize participation within a community scales with the ratio of population to assembly size, while aligning fractional community shares scales inversely with the number of available slack seats. These observations suggest convergence should be reasonably fast in practice.
\end{remark}

%% file: sections/upward-mobility.tex
\section{Upward Mobility}\label{section:pragmatics}

As in the previous work on this topic~\cite{halpern2024federated}, we would like the system to satisfy \emph{upward-mobility}, 
where members of a higher assembly are selected among members of the lower assembly.  Upward mobility has two goals:
\begin{enumerate}
    \item \textbf{Experience:}  Choose to higher assemblies seasoned members of lower assemblies.
    \item \textbf{Communication:} Enable vertical communication and the transfer of expertise and political agendas
    among higher and lower assemblies.
\end{enumerate}
In the static `one-shot' setting of~ \cite{halpern2024federated} there was only one design choice: Whether or not to require members of the higher assembly to be selected from among the members of the lower assemblies. 
Practically, doing so  allows members of a higher assembly to experience lower assembly work, and vice versa, in parallel.  This is less than ideal as the simultaneous serving in low- and high-level assemblies does not really provide for more seasoned members in high-level assemblies, since in this `one-shot' scenario they commence their two terms simultaneously. 
That work presented an algorithm that satisfied upward mobility,  the ex post, and the ex ante requirements simultaneously in the scenarios tested, but it was not proven to do so in general.

In the current dynamic setting the design-space for addressing upward mobility is much broader, and as we pioneer this direction we have no precedents to consult in choosing among the various design alternatives, only common sense. One attractive approach is the ``streaming'', or ``trickle up'', of representatives from lower assemblies to higher assemblies, by selecting a representative to a higher assembly as soon as they finish their term in the lower assembly. The sooner the better since the fresh experience and personal contacts made at the lower assembly can be brought to bear immediately at the higher assembly.  
We offer to take this approach as an extension to the protocol presented here: The next assembly member of a community is selected among the members of its population that best-satisfy the fairness requirements, with ties broken in favor of a most-recent veteran of a child assembly.  
As the pipeline of seasoned assembly members takes time to be built, and may be disrupted each time the federation structure changes, the best one can hope for is that eventually a higher assembly will consist of only seasoned members of the lower assembly, once the federation structure stabilizes.   

While highly-structured counterexamples show that achieving this precisely may not always be possible (with ``seasoned'' meaning retired less than one term ago; Daniel Halpern, personal communication), we expect simulations to show that once a federation stabilizes, and following a ``warm-up'' period, this tie-breaking rule almost always chooses members of a high-level assembly among recent (less than one term ago) veterans of the child assemblies. And we expect to prove, in future work, that in the limit there is a constant $c$ (possibly a function of the size of the assembly and the population)  after which suitable veterans can always be found after less than $c$ terms. 

The advantages of this approach is that it:
\begin{enumerate}
    \item Does not require overlapping membership in lower and higher assemblies.
    \item Feeds seasoned representatives from lower to higher assemblies.
    \item Enables vertical communication through personal contacts among lower and higher assembly members.
\end{enumerate}

%% file: sections/outlook.tex
\section{Outlook}\label{section:outlook}

We have presented a framework for grassroots federations together with a protocol that satisfies natural fairness requirements. This provides the means for decentralized communities to organize and govern themselves, at scale. Future research directions include:
\begin{itemize}
\item
\textbf{Simulations}: 
While our results show that fairness holds in the limit, simulations can shed further light on actual convergence speeds. We plan to quantify convergence under different federation structures and event dynamics.
\item
\textbf{Limitations}: 
We have made several simplifying assumptions, such as bounding the number of children per federation. Removing these is a natural direction; some are immediate (e.g., uniform assembly sizes, added only for simplicity---in practice, each federation could adopt its own constitution with its own assembly size).
Another direction is to study fairness from an online perspective: our current analysis reasons about eventual outcomes once the federation stabilizes, whereas an online view would assess fairness at finite horizons and compare GFP to an offline benchmark with full foresight.
\item 
\textbf{Grassroots proof}: A similar, yet simpler, `direct democracy' Grassroots Federation protocol was already shown to be grassroots~\cite{shapiro2025atomic,lewis2026volitional}. We expect the proof to carry over to the `assembly-based' protocol presented here; doing so is a subject of future work.

\item 
\textbf{Sybil-resilient grassroots federation}:
As digital communities are often vulnerable to Sybil attacks~\cite{douceur2002sybil}, our framework could in principle be extended with mechanisms providing protection against such attacks, and then applied to grassroots federations.
\end{itemize}

%% file: sections/acknowledgments.tex
\subsection*{Acknowledgments}

Co-funded by the European Union under the Horizon Europe project PERYCLES (Grant Agreement No. 101094862).

We thank Ariel Procaccia and Daniel Halpern for friendly and helpful discussions and feedback.